\documentstyle[preprint,prl,aps,epsf]{revtex}

\begin{document}

\title{
Further Search for the Decay $K^+ \to \pi^+ \nu \bar\nu$}

\author{S.~Adler$^1$, M.S.~Atiya$^1$, I-H.~Chiang$^1$, M.V.~Diwan$^1$,
  J.S.~Frank$^1$, J.S.~Haggerty$^1$, V.~Jain$^1$,
S.H.~Kettell$^1$, 
  T.F.~Kycia$^1$\cite{ted},
K.K.~Li$^1$, L.S.~Littenberg$^1$, C.~Ng$^1$\cite{chi},
  R.C.~Strand$^1$, C.~Witzig$^1$, M.~Kazumori$^2$,
T.K.~Komatsubara$^2$, M.~Kuriki$^2$,
  N.~Muramatsu$^2$, A.~Otomo$^2$,
S.~Sugimoto$^2$, T.~Inagaki$^3$, S.~Kabe$^3$,
  M.~Kobayashi$^3$, Y.~ Kuno$^3$, T.~Sato$^3$, T.~Shinkawa$^3$,
  Y.~Yoshimura$^3$, Y.~Kishi$^4$, T.~Nakano$^4$, T.~Sasaki$^4$,
  M.~Ardebili$^5$, A.O.~Bazarko$^5$, M.R.~Convery$^5$,
  M.M.~Ito$^5$, D.R.~Marlow$^5$,
  R.A.~McPherson$^5$, P.D.~Meyers$^5$, F.C.~Shoemaker$^5$,
  A.J.S.~Smith$^5$, J.R.~Stone$^5$, M.~Aoki$^6$\cite{masa},
  P.C.~Bergbusch$^6$, E.W.~Blackmore$^6$, D.A.~Bryman$^6$, 
A.~Konaka$^6$,
  J.A.~Macdonald$^6$, J.~Mildenberger$^6$, T.~Numao$^6$,
P.~Padley$^6$,   J.-M.~Poutissou$^6$,
R.~Poutissou$^6$,
  G.~Redlinger$^6$, P.~Kitching$^7$ and R.~Soluk$^7$\\
 (E787 Collaboration) }

\address{ $^1$Brookhaven National Laboratory, Upton, New York 11973\\
  $^2$High Energy Accelerator Research Organization (KEK),
  Tanashi-branch, Midoricho, Tanashi, Tokyo 188-8501, Japan \\ $^3$High
  Energy Accelerator Research Organization (KEK), Oho, Tsukuba,
  Ibaraki 305-0801, Japan \\ $^4$RCNP, Osaka University,
  10--1 Mihogasaki, Ibaraki, Osaka 567-0047, Japan \\ $^5$Joseph Henry
  Laboratories, Princeton University, Princeton, New Jersey 08544 \\
  $^6$ TRIUMF, 4004 Wesbrook Mall, Vancouver, British Columbia,
  Canada, V6T 2A3\\
$^7$ Centre for Subatomic Research, University of Alberta, Edmonton,
Alberta, T6G 2N5}

\maketitle

\begin{abstract}
A search for additional evidence for the rare kaon decay $K^+ \to \pi^+
\nu \bar\nu$ ~ has been made with a new data set comparable in
sensitivity to the previous exposure that produced a single event.
No new events were found in the pion momentum region examined,
$211<P<229$ MeV/$c$. Including a reanalysis of the original data set,
the backgrounds were estimated to contribute $0.08\pm0.02$ events.
Based on one observed event, the new
branching ratio
is $B$($K^+
\to \pi^+ \nu \bar\nu$)$=1.5^{+3.4}_{-1.2} \times 10^{-10}$.
\end{abstract}


\draft

Evidence for the decay $K^+ \to \pi^+ \nu \bar\nu$ ~
at a
branching ratio of $B$($K^+ \to \pi^+ \nu \bar\nu$)
$=4.2^{+9.7}_{-3.5}\times10^{-10}$
based on the
observation of a single event has been
reported by our group\cite{pnn95}.
 In the Standard Model (SM) calculation of
$B$($K^+ \to \pi^+ \nu \bar\nu$), the dominant  effects of the top quark
in second order weak loops make this flavor-changing neutral current
decay very sensitive to  $V_{td}$,
the coupling of the
top to down quarks in the Cabibbo-Kobayashi-Maskawa (CKM) quark mixing
matrix.
A fit based on the current phenomenology
gives a prediction of $(0.82 \pm 0.32)
\times 10^{-10}$ for this branching ratio
\cite{newbb}.  If constraints from
$|V_{ub}/V_{cb}|$ and $\epsilon_K$ are not imposed, a limit
$B(K^+ \to \pi^+ \nu\bar\nu) <
 1.67 \times 10^{-10}$
can be extracted that is almost entirely free of
theoretical uncertainties.
Although our initial observation is
consistent with the SM prediction,
the possibility of a larger-than-expected branching ratio\cite{lte}
gives further impetus for additional measurements.  In this paper, we
present results of a combined analysis of
the 1995 sample \cite{pnn95} and a new data sample of
comparable sensitivity.
All data were taken
with the E787 apparatus \cite{det}
at the Alternating Gradient Synchrotron (AGS)
of Brookhaven National Laboratory.

In the
decay $K^+ \! \rightarrow \! \pi^+ \nu \overline{\nu}$ at rest,
the $\pi^+$ momentum endpoint is
$227$ MeV/$c$.  Definitive
recognition of this signal requires that no other observable activity
is present in the detector
and all backgrounds are
suppressed
below the sensitivity for the signal.  Major background sources
include the two-body decays $K^+ \!  \rightarrow \! \mu^+ \nu_\mu$
($K_{\mu 2}$) and $K^+ \!  \rightarrow \!  \pi^+ \pi^0$ ($K_{\pi
2}$), scattered pions in the beam, and $K^+$ charge exchange
(CEX) reactions resulting in decays
$K_L^0\to\pi^+ l^- \overline{\nu}_l$,
where $l=e$ or $\mu$.
The E787 detector was designed to effectively distinguish these
backgrounds from the signal.

In the new data sets, taken during the 1996 and 1997 runs of the AGS,
kaons of about 700~MeV/$c$ were incident on the apparatus at a rate of
$(4 - 7) \times 10^6$ per 1.6-s spill.
The kaons were detected and identified by
\v{C}erenkov, tracking, and energy loss ($dE/dx$) counters.  About
25\% of the incident kaons reached an active target,
primarily consisting of 413 5-mm square scintillating fibers
which were used
for kaon and pion tracking.
Measurements of the momentum ($P$),
range ($R$) and kinetic energy ($E$)
of charged decay products were made using the target, a central drift
chamber, and a cylindrical range stack with 21 layers of plastic
scintillator and two layers of straw chambers (RSSC's),
all confined within
a 1-T solenoidal magnetic field.
The $\pi^+
\!  \rightarrow \! \mu^+ \!  \rightarrow \!  e^+$ decay sequence
of the decay products
in the range stack was observed using 500-MHz transient digitizers.
Photons
were detected in a $4\pi$-sr calorimeter consisting of a
14-radiation-length-thick barrel detector made of lead/scintillator
sandwich and 13.5-radiation-length-thick endcaps of
 undoped CsI crystals.
In comparison with Ref.~\cite{pnn95}, the
newer data were taken at a lower $K^+$ momentum
to reduce accidental hits\cite{rate},
and improvements were made to the 
trigger and data acquisition systems to take data
more efficiently.

The data were analyzed with the goal of reducing the total expected
background to significantly less than
one event in the combined data sample.
A decay particle was positively identified
as a $\pi^+$ using $P$, $R$ and $E$, and by the 
$\pi^+
\!  \rightarrow \! \mu^+ \!  \rightarrow \!  e^+$ decay sequence.
Events associated with any other decay products including photons
or with beam particles were efficiently eliminated by utilizing
the detector's full coverage of the $4\pi$ solid angle.
The requirements of a clean hit pattern in the target and
a delayed decay at least 2~ns after an identified $K^+$
suppressed background events due to CEX and scattered beam $\pi^+$.
In this work, the search for $K^+ \to \pi^+ \nu \bar\nu$
events was
restricted to the measured momentum region $211<P_{\pi^+}<229$ MeV/$c$,
between the
$K_{\mu 2}$ and $K_{\pi 2}$ peaks, to further limit backgrounds.

Compared to the analysis of Ref.~\cite{pnn95},
improvements in the kinematic reconstruction routines were made 
to reduce the tails of the $P$, $R$ and $E$
resolution functions.
In the new analysis, the position of the
incident kaon at the last beam counter
was used to aid in the determination of
the correct stopping position of the kaon in the target,
and thus to reduce the uncertainty in the
pion range masked by the kaon track.
Accidental hits that might have been
included in the pion energy measurements were identified and
removed more efficiently.
The measurement of the $z$ component
(the direction of the symmetry axis of the detector)
of the pion track in the range stack
was improved by using only the projection of the drift chamber fit and
the end-to-end timing in the range stack scintillators,
but excluding the $z$ information from the RSSC's, which had
a long resolution tail.
These changes resulted in some shifts
in the kinematic values found for
individual events, but the average quantities stayed the same.
In addition, improvements were made in the
particle identification criteria, particularly in the 
measurements of the $\pi^+ \rightarrow \mu^+ \rightarrow e^+$ decay
sequence.

Overall optimization
of the signal selection and background rejection criteria resulted in
roughly
a factor of two reduction of the expected backgrounds per kaon
decay  and
an increase of 25\% in the
acceptance for $K^+ \!  \rightarrow \! \pi^+ \nu \overline{\nu}$
in the 1995 sample.
 For the entire 1995--1997 exposure, the numbers of background events
expected from the sources mentioned above were $b_{{K_{\mu 2}}} = 0.03
\pm 0.01$, $b_{{K_{\pi 2}}} = 0.02 \pm 0.01$, $b_{Beam} = 0.02\pm
0.02$ and $b_{CEX} = 0.01 \pm 0.01$. In total, the background level
anticipated with the final analysis cuts was $b=0.08\pm 0.02$ events.
Tests of the background estimates near the signal region confirmed the
expectations.
The acceptance for $K^+ \to \pi^+ \nu \bar\nu$,
$A=0.0021 \pm 0.0001^{stat} \pm 0.0002^{syst}$, was derived from the
factors given in Table~\ref{acceptance} \cite{f_s}.
  The estimated systematic
uncertainty in the acceptance of about $10\%$
was due mostly to the uncertainty in pion-nucleus interactions.

Analysis of the full data set yielded only the single event previously
reported.
This result is shown in Fig.~1,
the range (in equivalent cm of scintillator)
vs. kinetic energy plot of events surviving all other cuts.
The revised kinematic values of the observed event
are $P=218.2 \pm 2.7$~MeV/$c$, $R=34.7 \pm 1.2$~cm
and $E=117.7 \pm 3.5$~MeV~\cite{pre}.
Based on one observed event,
the acceptance $A$ and the total exposure of
$N_{K^+}=3.2 \times 10^{12}$ kaons entering the target,
the new
value for the branching ratio is $B(K^+ \! \rightarrow \!  \pi^+ \nu
\overline{\nu}) = 1.5^{+3.4}_{-1.2} \times 10^{-10}$.

Using the relations given in Ref.\cite{newbb}
and varying each of the input parameters with the limits
given therein\cite{charm},
the present result provides a constraint,
$0.002 < |V_{td}| < 0.04$.
The extraction of these limits requires knowledge of
$V_{cb}$ and the assumption of CKM unitarity.  Alternatively, one can extract
corresponding limits on the quantity $|\lambda_t|$ ($\lambda_t \equiv
V^*_{ts} V_{td}$): $1.07 \times 10^{-4} < |\lambda_t| < 1.39 \times 10^{-3}$, 
without reference to the $B$-decay system.  In addition, the limits 
$-1.10 \times 10^{-3} <$ Re$(\lambda_t) < 1.39 \times 10^{-3}$ and
Im$(\lambda_t) < 1.22 \times 10^{-3}$ can be obtained
from our result\cite{qual}.
The latter is of particular interest because 
Im($\lambda_t$) is proportional
to the Jarlskog invariant\cite{jarls} and thus to
the area of the unitarity triangle.
Our result bounds this quantity without reference to the $B$-decay
system or to measurements of CP violation in $K_L^0 \to \pi\pi$ decays.

The limit found
in the search for decays of the form $K^+ \!  \rightarrow \!  \pi^+
X^0$, where $X^0$ is a neutral weakly interacting massless particle
\cite{x0},
is $B(K^+ \!  \rightarrow \!  \pi^+ X^0) < 1.1 \times 10^{-10}$ (90\%
CL), based on zero events observed in a $\pm 2 \sigma$ region around
the pion kinematic endpoint.

\acknowledgements

We gratefully acknowledge the dedicated effort of the technical
staff supporting this experiment and of the Brookhaven AGS
Department.
We thank A. J. Buras for useful discussions.
This research was supported in part by the
U.S. Department of Energy under Contracts No. DE-AC02-98CH10886,
W-7405-ENG-36, and grant DE-FG02-91ER40671, by the Ministry of
Education, Science, Sports and Culture of Japan
through the Japan-U.S. Cooperative Research Program
in High Energy Physics and under the Grant-in-Aids for
Scientific Research, for Encouragement of Young Scientists and for
JSPS Fellows,
and by the Natural
Sciences and Engineering Research Council and the National Research
Council of Canada.

\begin{table}
\begin{center}
\begin{tabular}{|l|l|}
~Acceptance factors& \\
\hline
~$K^+$ stop efficiency &   $0.704 $ \\
~$K^+$ decay after 2 ns&  $0.850 $ \\
~$K^+ \to \pi^+ \nu \bar\nu$~  phase space &   $ 0.155 $ \\
~Solid angle acceptance         &   $0.407 $\\
~$\pi^+$ nucl. int., decay-in-flight & $0.513$ \\
~Reconstruction efficiency &          $0.959 $ \\
~Other kinematic constraints &              $0.665 $ \\
~$\pi \rightarrow \mu \rightarrow e$ decay acceptance&     $ 0.306 $\\
~Beam and target  analysis &           $0.699 $\\
~Accidental loss &         $0.785 $\\
\hline
Total acceptance&   $0.0021 $\\
\end{tabular}
\end{center}
\caption{\label{acceptance}}{Acceptance factors used in the
measurement of $K^+ \to \pi^+ \nu \bar\nu$. The ``$K^+$ stop
efficiency'' is the fraction of kaons entering the target that
stopped \cite{f_s},
and ``Other kinematic constraints'' includes kinematic
particle identification and $dE/dx$ cuts.  }

\end{table}

\begin{figure}
\begin{center}
\centerline{\epsfbox{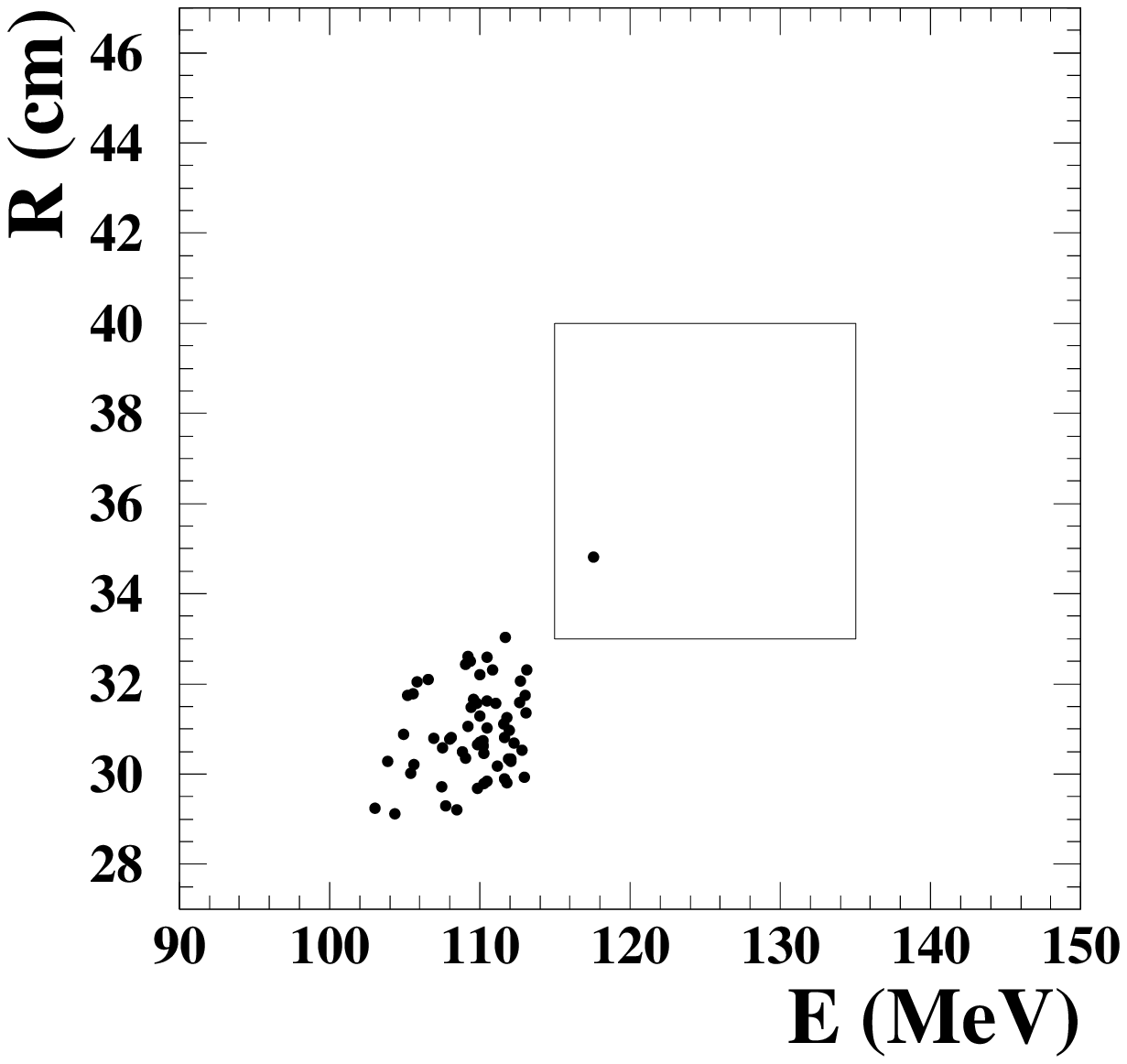}}
\end{center}
\vspace*{1cm}
Fig. 1.\hspace*{0.2cm}
Range (equivalent cm of scintillator) vs. Kinetic energy (MeV)  plot of
the final sample. The group of events around $E=108$~MeV
is due to the $K_{\pi 2}$ background.
The box indicates the acceptance region.
\end{figure}


\begin{references}
\bibitem{rate}At lower incident momentum
the fraction
of kaons stopping in the target increases
although the incident rate of kaons is reduced.
Since the rate of accidental hits
in the E787 detector,
which cause a loss of efficiency,
is proportional to the incident kaon flux,
there is net gain in the experimental efficiency
at the lower values of kaon momentum for comparable stopping rates.
\bibitem{pre}The previous values were $P=219.1$~MeV/$c$,
$R=36.3$~cm and $E=118.9$~MeV.
Although the new values are closer to the
$K_{\pi 2}$ background peak,
the probability that the event is due to this background remains very
low (0.003) because  the tails in the kinematic
quantities have been reduced by
the improved reconstruction routines described in the text.
\bibitem{charm}
The parameter values used in the calculations are:
$V_{cb} = 0.040 \pm 0.003$, $m_t =166 \pm 5$~GeV/$c^2$,
$M_W = 80.41$~GeV/$c^2$, $\lambda = 0.22$ and
 $P_0(X) = 0.42 \pm 0.06$ (see Ref.\cite{newbb} for definitions).
\bibitem{qual}
The limits on $\lambda_t$ all require knowledge of $m_t$.  Those on 
$|\lambda_t|$ and Re($\lambda_t$) also require knowledge of $m_c$ and
of the QCD scale, but are only modestly dependent on the values of these
quantities.
These quantities were varied over the ranges recommended
in Ref.\cite{newbb}.
\end{references}
\end{document}